\documentclass[sigconf,10pt]{acmart}

\usepackage{bbm}
\usepackage{verbatim}
\usepackage{graphicx}
\usepackage{url}
\usepackage{caption}
\usepackage{subcaption}
\usepackage{amssymb}
\usepackage{amsmath}
\usepackage{color}
\usepackage{algorithm, algorithmicx, algpseudocode}
\usepackage{siunitx}
\usepackage[printonlyused]{acronym}
\usepackage{cleveref}
\usepackage{lipsum}
\DeclareSIUnit{\belmilliwatt}{Bm}
\DeclareSIUnit{\dBm}{\deci\belmilliwatt}
\DeclareSIUnit{\isotropic}{Bi}
\DeclareSIUnit{\dBi}{\deci\isotropic}

\acrodef{mmW}{Millimeter-wave}
\acrodef{CPR}{compressive sensing phase retrieval}
\acrodef{BS}{base station}
\acrodef{UE}{user equipment}
\acrodef{SOTA}{state of the art}
\acrodef{AoA}{angle of arrival}
\acrodef{AoD}{angle of departure}
\acrodef{AWV}{antenna weight vector}
\acrodef{ADC}{analog-to-digital converter}
\acrodef{BB}{baseband}
\acrodef{RSRP}{reference signal received power}
\acrodef{CSI}{channel state information}
\acrodef{RSS-MP}{received signal strength matching pursuit}
\acrodef{COTS}{commercial-off-the-shelf}
\acrodef{PN}{pseudorandom noise}
\acrodef{NN}{neural network}
\acrodef{PAA}{phased antenna array}
\acrodef{CS}{compressive sensing}
\acrodef{LoS}{line-of-sight}
\acrodef{NLoS}{non-line-of-sight}
\acrodef{IA}{initial access}
\acrodef{DFT}{discrete Fourier transform}
\acrodef{AP}{access point}
\acrodef{MS}{mobile station}
\acrodef{RF}{radio frequency}
\acrodef{MPC}{multipath component}
\acrodef{BF}{beamforming}
\acrodef{SNR}{signal-to-noise ratio}
\acrodef{RSS}{received signal strength}
\acrodef{SINR}{signal-to-interference-plus-noise ratio}
\acrodef{DSP}{digital signal processing}
\acrodef{MIMO}{multiple-input multiple-output}
\acrodef{IC}{integrated circuits}
\acrodef{PS}{phase shifter}
\acrodef{DAC}{digital-to-analog converter}
\acrodef{EVM}{error vector magnitude}
\acrodef{Tx}{transmitter}
\acrodef{Rx}{receiver}
\acrodef{PCB}{printed circuit board}
\acrodef{EIRP}{effective isotropic radiated power}
\acrodef{ReLU}{rectified linear unit}
\acrodef{FC}{fully connected}

\newcommand{\tx}[0]{\text{T}}
\newcommand{\rx}[0]{\text{R}}
\newcommand{\hermitian}[0]{\text{H}}
\newcommand{\transpose}[0]{\text{T}}

\AtBeginDocument{%
  \providecommand\BibTeX{{%
    \normalfont B\kern-0.5em{\scshape i\kern-0.25em b}\kern-0.8em\TeX}}}

\setcopyright{acmcopyright}
\copyrightyear{2018}
\acmYear{2018}
\acmDOI{10.1145/1122445.1122456}

\acmConference[mmNet '20]{mmNet '20: the 4th ACM Workshop on Millimeter-Wave Networks and Sensing Systems}{September 25, 2020}{London, UK}
\acmBooktitle{mmNet '20: the 4th ACM Workshop on Millimeter-Wave Networks and Sensing Systems, September 25, 2020, London, UK}
\begin{document}

% \title{Implementation of Machine Learning assisted Noncoherent Compressive Millimeter-Wave Beam Alignment}
\title{mmRAPID: Machine Learning assisted Noncoherent Compressive Millimeter-Wave Beam Alignment}

\author{Han Yan}
\affiliation{%
  \institution{UCLA}}
%   \city{Los Angeles}
%   \country{United States}}
\email{yhaddint@ucla.edu}

\author{Benjamin W. Domae}
\affiliation{%
  \institution{UCLA}}
%   \city{Los Angeles}
%   \country{United States}}
\email{bdomae@ucla.edu}

\author{Danijela Cabric}
\affiliation{%
  \institution{UCLA}}
%   \city{Los Angeles}
%   \country{United States}
% }
\email{danijela@ee.ucla.edu}

% \author{Anonymous Authors}
\renewcommand{\shortauthors}{H. Yan, B. Domae, D. Cabric}
% \renewcommand{\shortauthors}{Anonymous Authors}

\begin{abstract}
Millimeter-wave communication has the potential to deliver orders of magnitude increases in mobile data rates. A key design challenge is to enable rapid beam alignment with phased arrays. Traditional millimeter-wave systems require a high beam alignment overhead, typically an exhaustive beam sweep, to find the beam direction with the highest beamforming gain. Compressive sensing is a promising framework to accelerate beam alignment. However, model mismatch from practical array hardware impairments poses a challenge to its implementation. In this work, we introduce a neural network assisted compressive beam alignment method that uses noncoherent received signal strength measured by a small number of pseudorandom sounding beams to infer the optimal beam steering direction. We experimentally showcase our proposed approach with a 60GHz 36-element phased array in a suburban line-of-sight environment. The results show that our approach achieves post alignment beamforming gain within 1dB margin compared to an exhaustive search with 90.2\% overhead reduction. Compared to purely model-based noncoherent compressive beam alignment, our method has 75\% overhead reduction. \end{abstract}

\maketitle
%%%%%%%%%%%%%%%%%%%%%%%%%%%%%%%%%%%
%
%        Introduction
%
%%%%%%%%%%%%%%%%%%%%%%%%%%%%%%%%%%%
\section{Introduction}
\ac{mmW} communication is a promising technology for future wireless networks, including 5G New Radio and 60 GHz Wi-Fi. Due to abundant spectrum, \ac{mmW} networks are expected to support ultra-fast data rates. As shown in both theory and prototypes, \ac{mmW} systems require \ac{BF} with large antenna arrays and narrow beams at both the \ac{Tx} and \ac{Rx} to combat severe propagation loss. Before data communication, directional beams must probe the channel to select a beam pair with adequate \ac{BF} gain. This procedure is referred as beam alignment\footnote{It is also referred as beam training, path identification, and path discovery.}. Existing \ac{mmW} systems use analog phased arrays with beam sweeping, an exhaustive search approach, for beam alignment. However, this method introduces high communication overhead. Further, the required number of channel measurements linearly scales with number of antenna elements, which is expected to increase with the evolution of mmW networks. 
% contributions
% \subsection{Contributions}
In this work, we present mmRAPID, \ac{mmW} Random Antenna weight vector based Path Identification without Dictionary. mmRAPID is a novel beam alignment method based on \ac{CS} theory, and it reduces the number of channel probings to logarithmically scale with antenna array size. We propose a machine learning approach to address a non-trivial \ac{CS} dictionary mismatch issue due to array hardware impairments. Our implementation and experiments using a \SI{60}{\giga\hertz} testbed demonstrated near perfect beam alignment with $90.2\%$ overhead reduction as compared to exhaustive beam sweeps. To the authors' best knowledge, this is the first work to experimentally demonstrate machine learning based beam alignment using a \SI{60}{\giga\hertz} phased array testbed and real measurement data. 

%%%%%%%%%%%%%%%%%%%%%%%%%%%%%%%%%%%%%%%%%%%%%%%%%%%%%%%%%%%%%%%%%%%%%%%%%%%%%%%

%paper organization

%%%%%%%%%%%%%%%%%%%%%%%%%%%%%%%%%%%%%%%%%%%%%%%%%%%%%%%%%%%%%%%%%%%%%%%%%%%%%%%
% \subsection{Organizations and notations}
The rest of the paper is organized as follows. \Cref{sec:literature} surveys \ac{mmW} fast beam alignment designs and proofs-of-concept. In \Cref{sec:system_model}, we present the problem statement and the motivation for using machine learning to solve it. The proposed design is presented in \Cref{sec:method}, followed by the implementation details with our \SI{60}{\giga\hertz} testbed in \Cref{sec:implementation}. The experimental results are presented in \Cref{sec:results}. Finally, \Cref{sec:Conclusion} concludes the paper.

%%%%%%%%%%%%%%%%%%%%%%%%%%%%%%%%%%%%%%%%%%%%%%%%%%%%%%%%%%%%%%%%%%%%%%%%%%%%%%%
%%%%%%%%%%%%%%%%%%%%%%%%%%%%%%%%%%%%%%%%%%%%%%%%%%%%%%%%%%%%%%%%%%%%%%%%%%%%%%%

% \textit{Notations:}
Scalars, vectors, and matrices are denoted by non-bold, bold lower-case, and bold upper-case letters, respectively.
% , e.g. $h$, $\mathbf{h}$ and $\mathbf{H}$.
The $(i,j)$-th element of $\mathbf{A}$ is denoted by $[\mathbf{A}]_{i,j}$. Similarly, the $i$-th element of a set $\mathcal{A}$ is denoted by $[\mathcal{A}]_{i}$. Transpose and Hermitian transpose are denoted by $(.)^{\transpose}$ and $(.)^{\hermitian}$ respectively. Inner product between $\mathbf{a}$ and $\mathbf{b}$ is denoted as $\langle \mathbf{a},\mathbf{b}\rangle$. $|\mathbf{a}|$ returns vector with magnitude of each element of $\mathbf{a}$.

%%%%%%%%%%%%%%%%%%%%%%%%%%%%%%%%%%%
%
%        Preliminary
%
%%%%%%%%%%%%%%%%%%%%%%%%%%%%%%%%%%%
\section{Related works}
\label{sec:literature}
% related works
Beam alignment for \ac{mmW} is an active research area. While some approaches focus on hardware innovations, e.g., fully-digital array and simultaneous frequency domain beam sweep facilitated by true-time-delay analog array \cite{boljanovic2020truetimedelay} or leaky wave antenna \cite{ghasempour2020single},
% , and novel analog front-end \cite{boljanovic2020design}.
others rely on signal processing. 

Model-based signal processing algorithms for beam alignment mainly rely on the sparsity of \ac{mmW} channels and a knowledge of the phased array response. State of the art approaches from this class of algorithms, namely hierarchical beam alignment and compressive sensing based beam alignment, have overheads that logarithmically scale with array size. The former uses sounding beams that adapt with previous measurement, bisecting the beam width to reduce the search space \cite{7885089}. The latter is based on either \ac{CS}, i.e., with coherent complex sample measurements, or \ac{CPR}, i.e., with noncoherent \ac{RSS} measurements~\cite{7400949,8777092,Rasekh_2018}. However, the model mismatch due to channel, antenna array and radio hardware impairments introduces non-trivial challenges.

%Model based signal processing mainly relies on the sparse \ac{mmW} channel and phased array model to design beam alignment algorithm. Two frameworks whose overheads logarithmic scale with array size are hierarchy beam alignment and compressive sensing based beam alignment. The former uses sounding beams that adapt with previous measurement in a bi-section manner to reduce search space \cite{7885089}. The latter is based on either \ac{CS}, i.e., with coherent complex sample measurements, or \ac{CPR}, i.e., with noncoherent \ac{RSS} measurements, that exploit sparsity of \ac{mmW} propagation \cite{7400949,8777092,Rasekh_2018}. However, the model mismatch which is due to both channel and radio hardware impairments introduces non-trivial challenges.

Data-driven signal processing can, from extensive training data, learn to infer the best beam using various low-overhead, in-band measurements and/or out-of-band information. In-band measurements include channel impulse response estimated by omni-directionally received pilots \cite{8395149} and a proportion of exhaustive beam search results \cite{9048770,8662770}. Out-of-band information includes the terminal's location \cite{8662770,9013296}. To date, these works either use a statistic channel model \cite{9048770} or ray tracing simulations \cite{8662770,9013296} to generate data.

With the increased availability of \ac{mmW} testbeds, there are many proofs-of-concepts. Work in \cite{kinget_CS_AOA} reports a chip-level demonstration of \ac{CS} based beam alignment with a channel emulator. \cite{Rasekh_2018,10.1145/3323679.3326532} showcase fast alignment by solving \ac{CPR} problems, while \cite{Hassanieh:2018:FMW:3230543.3230581,8878130} design and demonstrate fast beam alignment using multi-lobe sounding beams and combinatorics inspired algorithms. Work in \cite{8871119} reports experimental work that effectively reduces overhead when more than one spatial stream is used in a hybrid array. Finally, some prototypes also rely on the side information, e.g., sub-\SI{6}{\giga\hertz} \cite{7218630,10.1145/3117811.3117817} and visible light \cite{10.1145/3241539.3241542} measurements, for \ac{mmW} beam alignment.
%%%%%%%%%%%%%%%%%%%%%%%%%%%%%%%%%%%
%
%        Preliminary
%
%%%%%%%%%%%%%%%%%%%%%%%%%%%%%%%%%%%
\section{Noncoherent Compressive beam alignment}
%\section{\MakeLowercase{mm}RAPID Design}
\label{sec:system_model}
In this section, we start with the mathematical model and problem of noncoherent compressive beam alignment. As a reference, we also describe the state-of-the-art model-based solutions and their limitations.
%%%%%%%%%%%%%%%%%%%%%%%%%%%%%%%%%%%
%
%        System Model
%
%%%%%%%%%%%%%%%%%%%%%%%%%%%%%%%%%%%
\subsection{System model and problem statement}
We consider \ac{mmW} communication between an \ac{AP} 
\ac{Tx} and a \ac{MS} \ac{Rx}. The \ac{AP} and \ac{MS} are each equipped with an analog linear array with $N_{\tx}$ and $N_{\rx}$ elements. The channel follows an $L$-path geometric model $\mathbf{H} =\sum_{l=1}^{L}g_l \mathbf{a}_{\rx}(\phi_l)\mathbf{a}^{\hermitian}_{\tx}(\phi_l)$, where $\mathbf{a}_{\tx}(\theta) \in\mathbb{C}^{N_{\tx}}$,  $\mathbf{a}_{\rx}(\phi)\in\mathbb{C}^{N_{\rx}}$, and $g_l\in\mathbb{C}$ are the array responses in \ac{AP} and \ac{MS} and gain of the $l$-th path, respectively. Array responses are defined by their $n$-th element, i.e., $[\mathbf{a}_{\rx}(\phi)]_n = \mathrm{exp}(j2\pi (n-1)d/\lambda\sin(\phi))$ and $[\mathbf{a}_{\tx}(\theta)]_n = \mathrm{exp}(j2\pi (n-1)d/\lambda\sin(\theta))$, where $d$, $\lambda$, $\phi$ and $\theta$ are the element spacing, carrier wavelength, \ac{AoA} and angle of departure, respectively. We focus on the \ac{MS} \ac{Rx} side by assuming the the \ac{AP} \ac{Tx} \ac{AWV} $\mathbf{v} = \mathbf{a}_{\tx}(\theta_1)/\sqrt{N_{\rx}}$ is pre-designed. Thus, the channel model in the rest of the paper is
\begin{align}
    \mathbf{h} =\sum_{l=1}^{L}\alpha_l \mathbf{a}_{\rx}(\phi_l) \approx \alpha_1\mathbf{a}_{\rx}(\phi_1)\triangleq \alpha\mathbf{a}_{\rx}(\phi^{\star}),
    \label{eq:channel}
\end{align}
where $\alpha_l = \alpha_l\mathbf{a}^{\hermitian}_{\tx}(\theta_l)\mathbf{v}$ is the post-Tx-beam channel gain. 
The approximation in (\ref{eq:channel}) is from the selection of the \ac{Tx} beam, which results in $|\alpha_1|\gg |\alpha_l|,l>1$. We define $\phi^{\star}$ as the true \ac{AoA} of the channel. When the \ac{Rx} uses \ac{AWV} $\mathbf{w}$, the received symbol is 
\begin{align}
y = \mathbf{w}^{\hermitian}\mathbf{h}s + n,
\label{eq:symbol_measurement}
\end{align}
where $s$ is the \ac{Tx} symbol and $n$ is the post-combining thermal noise which is modeled as additive white Gaussian noise with variance $\sigma_{\text{n}}^2$. Without loss of generality, we let $s=1$ and define \ac{SNR} as $\mathrm{SNR} = |\alpha|^2/\sigma_{\text{n}}^2$.

We consider a codebook based communication protocol which consists of two phases: a beam alignment phase and a data communication phase. The channel is unknown to the \ac{AP} \ac{Rx} but can be assumed invariant between these two phases. During beam alignment, the \ac{MS} \ac{Rx} uses a sounding codebook, $\mathcal{W}_{\text{S}}$ (with $|\mathcal{W}_{\text{S}}|=M$ codewords), to probe the channel. The associated measurements are processed to select the best beam from a fixed directional codebook $\mathcal{W}_{\text{D}}$ (with $|\mathcal{W}_{\text{D}}|=K$ codewords), which is then used in the data communication. Each codeword of directional codebook is a steering vector, i.e., $[\mathcal{W}_{\text{D}}]_k = \mathbf{a}_{\rx}(\theta_k)/\sqrt{N_{\rx}}$, and these directions $\{\theta_k\}_{k=1}^{K}$ cover an angular region of interest.

Three additional assumptions are relevant to our implementation. 
Firstly, the sounding codebook $\mathcal{W}_{\text{S}}$ is loaded into hardware in advance and each codeword is applied in a sequential manner. Adaption that uses on-the-fly measurements to change either the codebook or the codeword selection order, e.g., a hierarchy search, is not desired. In fact, we focus on pseudo-random sounding codebooks $\mathcal{W}_{\text{S}}$, a well adopted design from compressive sensing literature when the \ac{Rx} does not have prior knowledge of the channel \cite{7400949,8777092,Rasekh_noncoherentCS_ACM_2017,Rasekh_2018}. Specifically, the magnitude of each \ac{AWV} in $\mathcal{W}_{\text{S}}$ is $1/\sqrt{N_{\rx}}$ and the phase is randomly picked from the set $\{0,\pi/2,\pi,3\pi/2\}$. These are referred to as \ac{PN} \ac{AWV} or beams in the remainder of the paper.
Secondly, the received symbols in (\ref{eq:symbol_measurement}) are not directly observable. Instead, each channel measurement is noncoherently taken from the preamble, a sequence of pilot symbols, in the form of \ac{RSS}. 
Lastly, the phased array is non-ideal and has realistic hardware impairments. An optimistic assumption is to model these impairments as gain and phase offset, i.e., an unknown multiplicative error $\mathbf{e}\in\mathbb{C}^{N_{\rx}}$ independent of the codewords \cite{10.1145/3323679.3326532}. With the above assumptions, the $M$ channel probings give \ac{RSS} $\mathbf{p} = [p_1,\cdots,p_M]^{\transpose}$ where the $m$-th probing is
\begin{align}
p_m = |\tilde{\mathbf{w}}_m^{\hermitian}\mathbf{h}|+n_m.
\label{eq:pRx_measurement}
\end{align}
In the equation, $\tilde{\mathbf{w}}_m = \mathrm{diag}(\mathbf{e})[\mathcal{W}_{\text{S}}]_m$ is the receiver combiner with hardware impairment, and $n_m$ is the error in RSS measurement. To that end, the compressive noncoherent beam alignment problem is:

\textit{\textbf{Problem:}} Design a signal processing algorithm that uses the noncoherent measurements $\mathbf{p}$ from (\ref{eq:pRx_measurement}) to infer the best directional beam for data communication, i.e., $\hat{\mathbf{w}}\in\mathcal{W}_{\text{D}}$.

The performance metrics are the required number of measurements $M$ and the post-alignment \ac{BF} gain, i.e., normalized gain in data communication phase $G = |\mathbf{h}^{\hermitian}\mathrm{diag}(\mathbf{e})\hat{\mathbf{w}}|^2/\|\mathbf{h}\|^2$. Note that an exhaustive search uses the same codebook for both alignment and communication, i.e., $\mathcal{W}_{\text{S}} = \mathcal{W}_{\text{D}}$. It is straightforward to find the optimal codeword $\mathbf{w}^{\star} = \max_{\mathbf{w}\in\mathcal{W}_{\text{D}}}G$ with overhead cost $M=K$. The goal is to reduce $M$ while introducing marginal impact to post-alignment \ac{BF} gain as compared to an exhaustive search\footnote{With increased codebook size $K$, beam steering with AWV $\mathbf{w}^{\star}$ is asymptotically the same as the steering towards ground truth \ac{AoA} $\phi_{\star}$. Thus, we do not directly compare with the latter in this work.}, e.g., <2dB loss. we further defined overhead reduction ratio as $(K-M)/K$.

%%%%%%%%%%%%%%%%%%%%%%%%%%%%%%%%%%%
%
%        Limitation of CS
%
%%%%%%%%%%%%%%%%%%%%%%%%%%%%%%%%%%%
\subsection{Model based solution and its limitation}
\label{sec:CPR_limitation}
% CS is a powerful signal processing tool. When directly deal with (\ref{eq:symbol_measurement}), i.e., collecting measurement $\mathbf{y} =[y_1,\cdots,y_M]$, 
The beam alignment with (\ref{eq:pRx_measurement}) can be formulated as a \ac{CPR} problem when the error $\mathbf{e}$ is assumed to be negligible, i.e.,
\begin{align}
    \mathbf{p} = |\mathbf{W}^{\hermitian}\mathbf{h}| + \mathbf{n} = |\mathbf{W}^{\hermitian}\mathbf{A}_{\rx}\mathbf{g}| + \mathbf{n} \triangleq |\boldsymbol{\Psi}\mathbf{g}| + \mathbf{n}.
    \label{eq:CPR}
\end{align}
In the above equation, $[\mathbf{A}_{\rx}]_k = \mathbf{a}_{\rx}(\theta_k)$, $\theta_k$ are the steering directions in the \ac{DFT} codebook (the \ac{AoA} hypothesis), and $\mathbf{g}\in\mathbb{C}^{K}$ is a sparse vector with all-zero elements except the $k^{\star}$-th being $\alpha$ (i.e. associated with true \ac{AoA} $k^{\star} = \{k|\theta_k = \phi^{\star}\}$). The error vector is $\mathbf{n} = [n_1,\cdots,n_M]^{\transpose}$. The sensing matrix $\mathbf{W}$ is from error-free sounding \ac{AWV}s and defined as $[\mathbf{W}]_m\triangleq[\mathcal{W}_{\text{S}}]_m$. The solution to a general \ac{CPR} model is guaranteed with an adequate number of measurements $M$, which linearly scales with the sparsity level, i.e., non-zero elements in $\mathbf{g}$, and logarithmically scales with $K$ \cite{6998861}. Solutions to a general \ac{CPR} can use convex optimizations \cite{Rasekh_2018,10.1145/3323679.3326532} or approximate massage passing \cite{6998861}. Solving \ac{CPR} in this work, or finding the sparsity level 1 vector $\mathbf{g}$ from $\mathbf{p}$, directly leads to a solution of beam alignment since the non-zero element, say $\hat{k}$-th, can simply be used to select the best beam from the \ac{DFT} codebook $\hat{\mathbf{w}} = [\mathcal{W}_{\text{D}}]_{\hat{k}}$. In this special case, the heuristic approach of \ac{RSS-MP} \cite{Rasekh_noncoherentCS_ACM_2017} also applies, where $\hat{k} = \text{arg}\max_{k} \langle\mathbf{p},|[\boldsymbol{\Psi}]_k|\rangle/\|[\boldsymbol{\Psi}]_k\|$.
% In other words, the received signal power is aligned into vector $\mathbf{p} = [p_1,\cdots, p_M]$. The algorithm uses $k^{\star} = \text{arg}\max_{k} \langle\mathbf{p},[\tilde{\boldsymbol{\Psi}}]_k\rangle/\|\tilde{\boldsymbol{\Psi}}_k\|^2$ where modified dictionary $\tilde{\boldsymbol{\Psi}}$ takes absolute value of original dictionary as $[\tilde{\boldsymbol{\Psi}}]_{m,n} = |[\boldsymbol{\Psi}]_{m,n}|$. The algorithm uses best directional beam to be $\mathbf{w}^{\star} = \mathbf{a}(\phi_{k^{\star}})$.

The key concern in existing \ac{CPR} solutions is the required knowledge of dictionary $\boldsymbol{\Psi}$ (aka magnitude and phase response of AWVs and beam patterns), in (\ref{eq:CPR}). With hardware impairments, the sensing matrix $\mathbf{W}$ in (\ref{eq:CPR}) is composed of distorted sounding \ac{AWV}s $[\mathbf{W}]_m=\mathrm{diag}(\mathbf{e})\mathbf{w}_m$. This can be problematic in a practical radio for three reasons.
Firstly, in the production of radio hardware, the error $\mathbf{e}$ are due to the combined effects of a systematic offset among all devices and a device dependent random offset \cite{9052488}. To date, it is generally cost-effective to only calibrate and compensate the common offset. Over-the-air calibration of device-dependent offsets is prohibitively expensive and time consuming. 
Secondly, the mainlobe of \ac{DFT} pencil beam is not sensitive to array offset \cite{9052488}. Although sidelobes are more vulnerable to distortion, they are not directly used in \ac{mmW} systems during beam sweeping or beam steering. Thus, leaving a device-dependent array offset can be acceptable.
%
% During calibration, the 1-st (as reference w.o.l.g) and the $n$-th antenna element are both turned-on and commanded to set to zero phase. A reference transmitter is placed at the bore-sight. Therefore, any phase offset among these two elements results in non-matched phase and thus a reduced received power. A trial and error tuning of the  their relative phase. There are two reasons that leads to a large $\psi_{\text{max}}$. Firstly, 
Lastly, the beam patterns of \ac{PN} \ac{AWV}s are sensitive to array offset, as will be shown in \Cref{sec:results}. Thus, \ac{CPR} based algorithms are likely to experience model mismatch and degraded performance. 
% We quantitatively study this degradation in \Cref{sec:pettern_distortion_sim} based on the error model. In actual radios, modeling $\mathbf{e}$ is more challenging, as they can be associated with other array-inherent defects. 
% In \Cref{sec:results} we perform an empirical evaluation using our testbed to verify the above argument.

%%%%%%%%%%%%%%%%%%%%%%%%%%%%%%%%%%%
%
%        Proposed Method
%
%%%%%%%%%%%%%%%%%%%%%%%%%%%%%%%%%%%
%\section{Neural network assisted compressive beam alignment}
% \section{\MakeLowercase{mm}RAPID Design}
\section{mmRAPID Design}

\label{sec:method}
To address the model mismatch in solving \ac{CPR}, we propose a data driven approach for beam alignment.

\subsection{Main idea of mmRAPID}
\label{sec:main_idea}
The key insight of our approach is that, although analytically solving the noncoherent beam alignment problem using model (\ref{eq:CPR}) is difficult, its solution can be easily found by an exhaustive beam sweep. Therefore, we can resort to a data driven approach to learn how to solve the \ac{CPR} problem (\ref{eq:CPR}) with unknown deterministic offset. The proposed system contains two stages, each covering a much longer time scale than the beam alignment or communication phases. 

We refer the first stage as the \textit{learning stage}, where the radio uses a concatenated codebook $\mathcal{W} = \mathcal{W}_{\text{D}} \cup\mathcal{W}_{\text{S}}$ for multiple beam alignment phases. Specifically, the sounding results from exhaustive search $\mathcal{W}_{\text{D}}$ provide the solution to the beam alignment problem; the so called \textit{labels} in machine learning terminology. The sounding results from $\mathcal{W}_{\text{S}}$ are treated as the \textit{features}, whose statistical relationship with the \textit{labels} can be extracted by machine learning tools, e.g., \ac{NN} or support vector machine. Admittedly, the beam alignment overhead in this stage is $K+M$, even higher than the overhead $K$ from an exhaustive search. The beam alignment \textit{features} and \textit{labels} must be collected in various environments to reliably generalize their relationship. In a practical system, this would arise from randomness in physical position and orientation of the \ac{MS}, e.g., a phone held by a human with different posture in different places. In fact, the learning stage can be completely ambient and does not require dedicated interaction from the user \cite{8395149}. 

We refer the second stage as the \textit{operation stage}, where \ac{MS} only uses codebook $\mathcal{W}_{\text{S}}$ for beam alignment, compressing the overhead. The algorithm then only uses the \textit{feature} to predict the \textit{label}, i.e., the best beam $\mathbf{w}^{\star}$ in data communication phase.

\subsection{Neural network design}
\label{sec:neuralnet}
In this work, we designed a dense \ac{NN} to predict the optimal \ac{DFT} beam for a given unknown channel $\mathbf{h}$, i.e., \textit{label}, based on \ac{PN} beam \ac{RSS} measurements, i.e., \textit{feature}. The network used 3 \ac{FC} layers, each using \ac{ReLU} activation functions. For all tested values of $M$, we used the same network architecture, with 64, 128, and $K$ units in the first through third \ac{FC} layers respectively.

Input \ac{RSS} data was expressed in linear scale and normalized by the maximum value. These feature transformations limit the data to the range $[0, 1]$, prevent activation function saturation, and improve the learning performance. Batch normalization layers were also used just before the \ac{ReLU} activations in first and second \ac{FC} layers as feature regularization to improve training efficiency.  

Our design used a sparse categorical cross entropy loss function to produce our classification results over the $K$ possible \ac{DFT} beam physical angle labels. For training, we used the $\mathrm{RMSprop}$ optimizer.  The network architecture was implemented and trained in Keras/Tensorflow. The total number of trainable parameters in this network depends on the input feature dimension ($M$) and the label dimension ($K$): $64M + 129K + 8768$.

% \color{red} Ben, please add more details for
% 1) What is the input. what's its dimension? 2) What is the data and label. Is it classification or regression problem. 3) What is the parameter in each layers. What is the total number of layer to train. 4) Performance and training metric, e.g., accuracy in classification problem or MSE in regression problem. 5) Different \ac{NN} architectures and design parameters (optional), e.g., dense layer only v.s. deep convolution model.

% The architecture is shown in \Cref{fig:NN_architecture}.
% \color{black}

% \begin{figure}
% \begin{center}
% \includegraphics[width=0.3\textwidth]{figures/mlbeamtraining_fcnet.png}
% \end{center}
% \vspace{-3mm}
% \caption{\color{red} Neural network figure. Need revision; Consider remove it if running out of space \color{black}}
% \vspace{-3mm}
% \label{fig:NN_architecture}
% \end{figure}

%%%%%%%%%%%%%%%%%%%%%%%%%%%%%%%%%%%
%
%        Implementation
%
%%%%%%%%%%%%%%%%%%%%%%%%%%%%%%%%%%%
\section{Implementation in mmW testbed}
\label{sec:implementation}
This section starts with a description of the testbed, followed by the \ac{NN} based beam alignment implementation.
%%%%%%%%%%%%%%%%%%%%%%%%%%%%%%%%%%%
%
%        Capability of HW and SW
%
%%%%%%%%%%%%%%%%%%%%%%%%%%%%%%%%%%%
% \subsection{Millimeter-wave testbed}
% \vspace{-1mm}
% \begin{table}
% \caption{Summary of mmW testbed}
% \centering
% \begin{tabular}{|l|l|}
% \hline 
% \textbf{Parameters} & \textbf{Values}\tabularnewline
% \hline 
% \hline 
% RF channel & IEEE802.11ad (\SI{60.48}{\giga\hertz}) \tabularnewline
% \hline
% Array geometry & Planar array (36 by 8 elements) \tabularnewline
% \hline
% Array control (az) & Individual phase $\&$ on-off control \tabularnewline
% Array control (el) & Fixed beam (bore-sight)\tabularnewline
% \hline
% Scan region (az) & \SI{-45}{\degree} to \SI{45}{\degree}\tabularnewline
% \hline
% Beam-width (az) & \SI{2.8}{\degree} (with directional codebook)\tabularnewline
% \hline 
% Max EIRP & \SI{45}{\dBm}\tabularnewline
% \hline 
% \end{tabular}
% \vspace{-4mm}
% \label{tab:sounder_parameter}
% \end{table}

Our testbed is the Facebook Terragraph (TG) channel sounder, a pair of TG nodes customized for measurements of \SI{60}{\giga\hertz} channels \cite{10.1145/3349624.3356763}. 
%
% Although the testbed is designed for modeling the channel characteristics such as the path loss, angular-power profile, and delay-power profile, e.g., \cite{10.1145/3349624.3356763}, its capability is adequate for beam alignment experiment. 
Each TG node has a 36 by 8 planar phased array. When using pencil beams, the narrowest one-sided \SI{3}{\decibel} beamwidth is \SI{1.4}{\degree} in the azimuth plane. Readers are referred to \cite{10.1145/3349624.3356763} for more details of the testbed. 
% The key hardware specifications of the testbed are listed in
% \Cref{tab:sounder_parameter}. 
% A default directional beam codebook with $K=64$ steering directions (with step-size $1.4^{\circ}$) between $-45^\circ$ to $45^\circ$ is provided.  
%
\begin{figure}
\begin{center}
\includegraphics[width=0.48\textwidth]{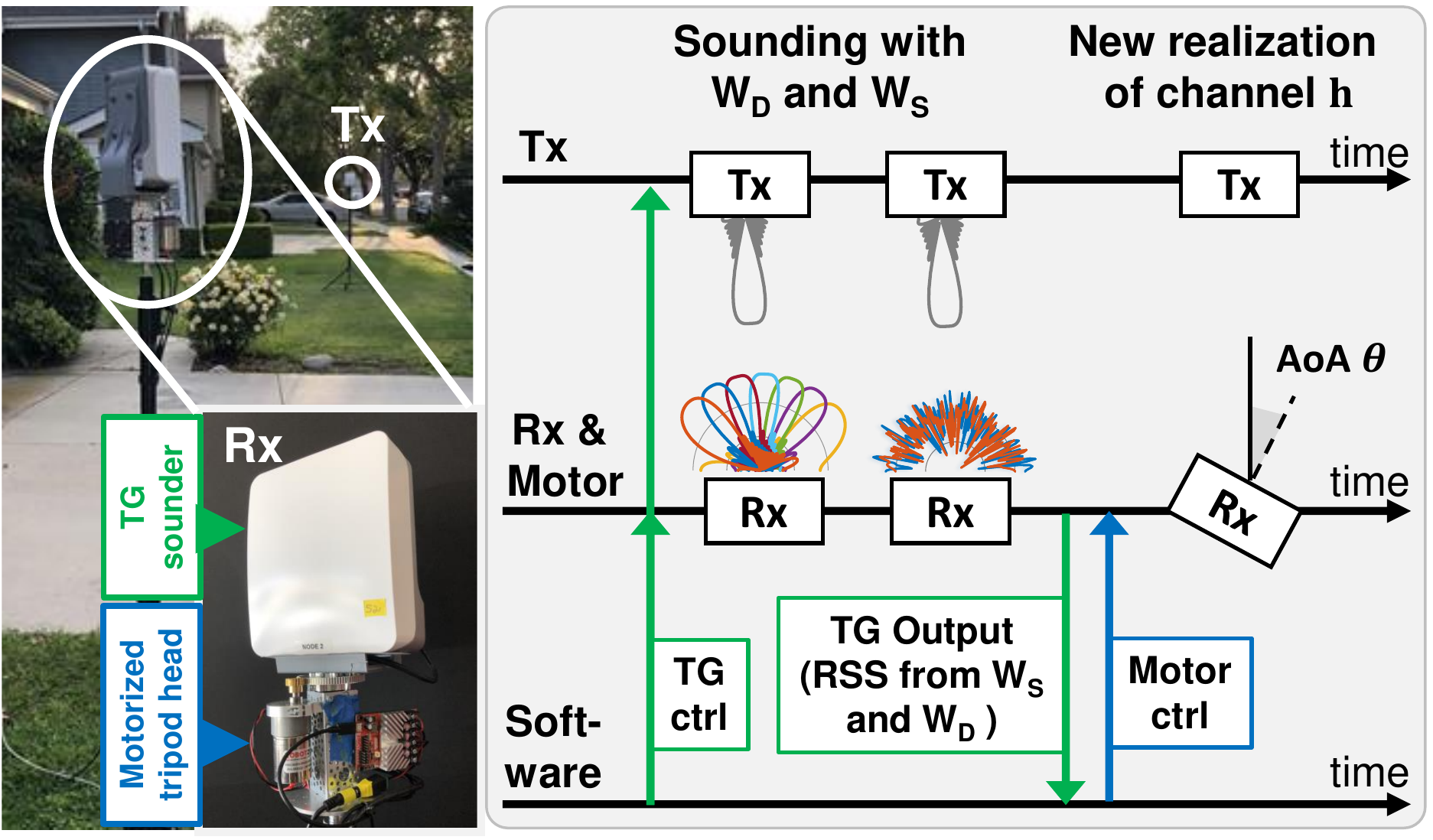}
\end{center}
\vspace{-4mm}
\caption{Overview of the testbed, experiment environment, and data capture procedure.}
% \caption{Overview of the experimental testbed. (a) shows the testbed environment. (b) depicts a TG radio attached to our motorized tripod head for automated data captures. (c) is a timeline of how the testbed captures each data point, with this process repeated thousands of times during the experiment.}
\vspace{-5mm}
\label{fig:testbed}
\end{figure}

The testbed has an application programmable interface (API) that allows a host computer to customize AWVs when transmitting or receiving IEEE802.11ad packets. The API also provides measurements from the received preamble, e.g., received signal strength indicator, short training field (STF) specified \ac{SNR}, and long-training field specific channel impulse response estimation. 
%
% The read outs includes beam indices of pre-loaded codebook, in-chip measures received signal strength indicator, \ac{SNR}, channel impulse response based on the preamble of received packet. 
Note that although multiple automatic gain control (AGC) amplifiers are involved in preamble measurements, a look-up table is used to make the impact of AGC transparent, which indicates the fidelity of model (\ref{eq:pRx_measurement}). The array offset was calibrated and compensated using a golden design instead of device-by-device calibration.

%%%%%%%%%%%%%%%%%%%%%%%%%%%%%%%%%%%
%
%        Implementation Details
%
%%%%%%%%%%%%%%%%%%%%%%%%%%%%%%%%%%%
\subsection{Data Capture Automation}
As mentioned in \Cref{sec:method}, it is desired for the \ac{NN} to learn from beam alignment data collected in various of locations and physical orientations of radios. To automate this procedure, we used a programmable motor on the receiver side. We designed a 3D printed bracket to attach the receiver to a motorized turntable kit controlled by a motor controller, as shown in \Cref{fig:testbed}. We mechanically rotated the receiver between \SI{-45}{\degree} and \SI{45}{\degree} from the transmitter boresight. The procedure achieved pseudo random realizations of \ac{AoA} of propagation channel $\mathbf{h}$ using pre-programmed motor positions. \Cref{fig:testbed} demonstrates how the receiver collected different physical AoAs using the automated turntable. Note that the motor was not precisely controlled, nor did it provide the true \ac{AoA}, unlike a turntable required for chamber calibration of the array. The motor's only purpose was to emulate random physical positions and hold the posture of the \ac{MS} as described in \Cref{sec:main_idea}. No such motor is required when generalizing this approach to actual scenarios.

%(Servocity Gear Drive Pan Kit)%
%(Basicmicro Motion Control Roboclaw 2x30A)%

% \color{red} Ben, please elaborate with 
% 1) The model of motor and control kit (including references). 2) The capability of motor. 3) How is motor connected with TG.\color{black}
%%%%%%%%%%%%%%%%%%%%%%%%%%%%%%%%%%%
%
%        Proposed Method
%
%%%%%%%%%%%%%%%%%%%%%%%%%%%%%%%%%%%
\section{Experiment and Results}
\label{sec:results}
This section describes the experiment details, followed by experimental results and comparison with state-of-the-art.

\subsection{Experiment details}
We conducted the experiment in a \ac{LoS}, suburban outdoor environment, shown in \Cref{fig:testbed}a. The radios were mounted on tripods 
% approximately \SI{2}{\meter} above the ground and 
separated by approximately \SI{14}{\meter} (\SI{91}{\decibel} pathloss). 
% with no obstructions directly neighboring the path. 
% The pathloss is near  from both Friis equation and post-processing of experiment data.
The azimuth \ac{AoA} of the channel is randomly changed by the motor before each capture.
% , the elevation angle was perfectly aligned and never changed. Similarly,
\ac{Tx} directional beam was pointed towards the \ac{Rx} at all time.

During each collection period, we collected 2,000 points. Each point consisted of 100 \ac{RSS} measurements using the sounding codebook $\mathcal{W}$, i.e., $K=64$ \ac{DFT} beams between \SI{-45}{\degree} and \SI{45}{\degree} such that adjacent pencil beams overlap by a half of one-sided \SI{3}{\decibel} beamwidth and $M_0=36$ \ac{PN} beams. Each data point spends \SI{9}{\second}, including \SI{7}{\second} when the Rx was static\footnote{The latency of testbed API is not optimized. Hence, our goal is to achieve alignment with a compressed number of probings $M$ instead of high speed.} and \SI{2}{\second} for the motor movement to create a new \ac{LoS} propagation direction. Although we collected data for $M_0=36$ \ac{PN} beams, only the first $M$ beams were used for training and testing with compressive beam alignment algorithms. A total of 3 collection periods from three different days were included in this paper's results, each with a different \ac{SNR}\footnote{Such SNR is measured by STF which has consistent definition with the $\mathrm{SNR}$ for (\ref{eq:symbol_measurement}). Although SNR in (\ref{eq:pRx_measurement}) cannot be directly measured, it should be larger than the STF-SNR since RSS measurements average a sequence of samples.}. We changed the \ac{SNR} by modifying the transmit \ac{EIRP} which leads to median PN beam SNRs of 10, 10, and 12 dB. After data collection, \textit{labels} with insufficient training data (at least 20 points per \ac{SNR}) were eliminated, leaving $K=51$ remaining labels, whose associated \ac{AoA} are between \SI{-26.4}{\degree} to \SI{43.6}{\degree}. Of the data associated with the $K=51$ \ac{DFT} beam labels, a total of 3,060 data points were used for training and 1898 points were used for evaluation. Note that all data will be made public upon publication. During evaluation, the \ac{DFT} sounding results were used as the ground truth to measure beam prediction performance from compressive PN probing.

% \color{red}
% Ben, please contribute with 1) describe environment and experiment details using \Cref{fig:environment}; Elevation beam is perfectly aligned. 2) what is the propagation condition. What is distance b/w \ac{Tx} and Rx. 3) how much data size is collected. How long does data capture take. what's their \ac{SNR} in terms of STF-SNR 4) behavior of transmitter. What beams are used. How many beams are used.
% 5) Behavior of receiver. What beams are used. How many beams are used. 6) Some other MISC like control angles within range. 7) When randomize training and testing work. 8) We always use sounding from the first M PN beams when sweeping M.
% \color{black}

For fair comparison, the training data was also available to the solution that analytically solves \ac{CPR}. Note that when using $[\boldsymbol{\Psi}]_k$ in (\ref{eq:CPR}) as collected labels of training data, the system can estimate $|\tilde{\mathbf{w}}^{\hermitian}_m\mathbf{a}_{\rx}(\theta_k)|$, i.e., the magnitude of dictionary $\boldsymbol{\Psi}_k$. Although such estimates cannot be directly used in \ac{CPR} as phase information of $\boldsymbol{\Psi}$ are missing, they help the \ac{RSS-MP} algorithm \cite{Rasekh_noncoherentCS_ACM_2017}. Hence, we refer to the vanilla \ac{RSS-MP} as one that uses only the model predicted dictionary, and dictionary refined \ac{RSS-MP} as the one enhanced by training data. 
Also, our best efforts in applying the convex optimization based \ac{CPR} solution led to unsatisfactory performance, likely due to the imperfect knowledge of the dictionary $\boldsymbol{\Psi}$ when using the PN sounding beam, similar to the finding from \cite{10.1145/3323679.3326532}.

% \begin{figure}
% \begin{center}
% \includegraphics[width=0.25\textwidth]{figures/environment.png}
% \end{center}
% \vspace{-2mm}
% \caption{\color{red} Environment figure. Need to be small but also need to include details. \color{black}}
% \vspace{-4mm}
% \label{fig:environment}
% \end{figure}
%%%%%%%%%%%%%%%%%%%%%%%%%%%%%%%%%%%
%
%        Conclusions
%
%%%%%%%%%%%%%%%%%%%%%%%%%%%%%%%%%%%
\subsection{Experiment results}

\begin{figure}
\begin{center}
\includegraphics[width=0.5\textwidth]{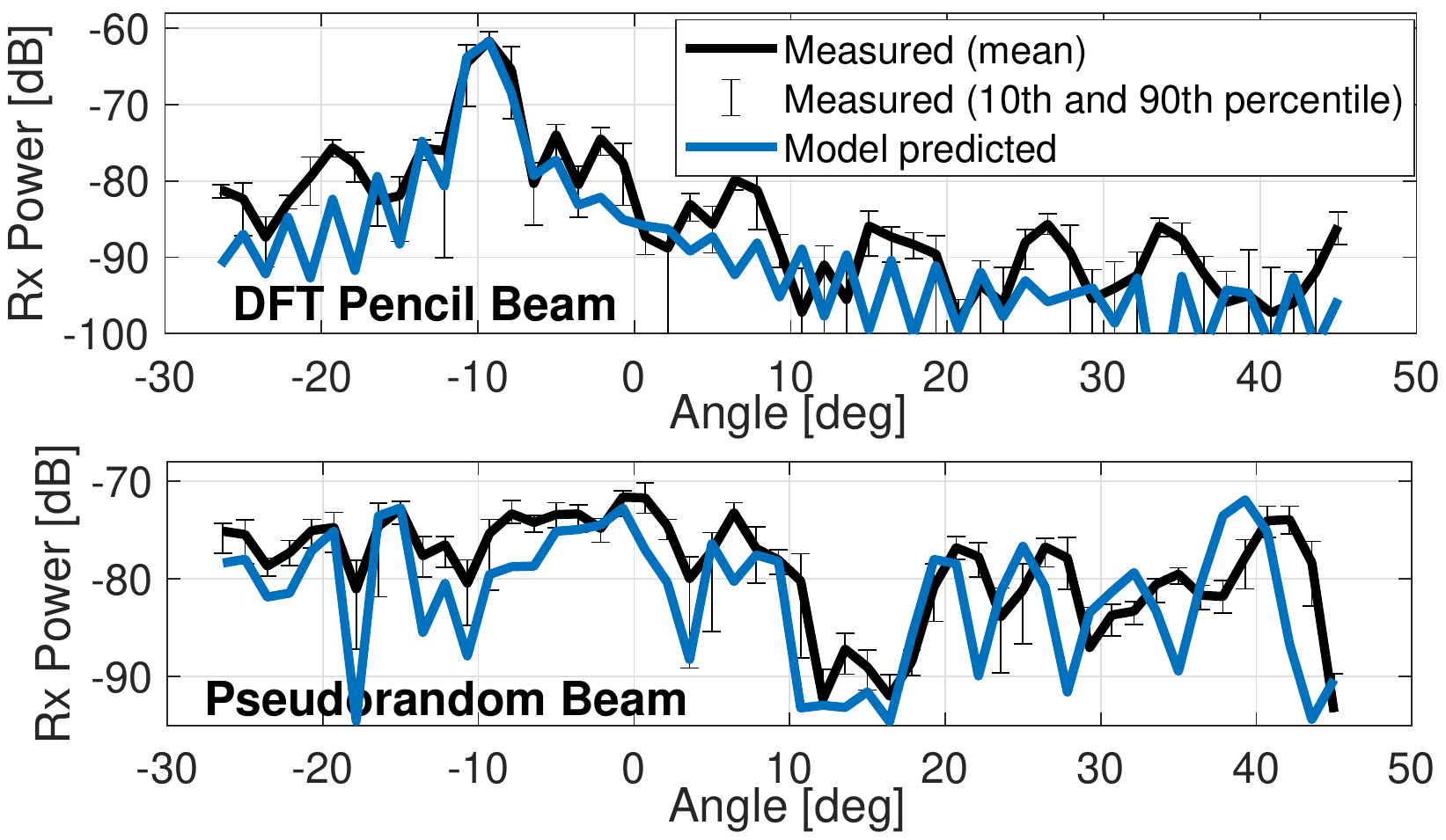}
\end{center}
\vspace{-3mm}
\caption{Measured and model predicted beam pattern of a \ac{DFT} beam $\mathbf{w}\in\mathcal{W}_{\text{D}}$ and a PN beam $\mathbf{w}\in\mathcal{W}_{\text{S}}$.}
\vspace{-3mm}
\label{fig:result0}
\end{figure}

The captured data allowed us to coarsely evaluate the beam pattern, $|\tilde{\mathbf{w}}^{\hermitian}\mathbf{a}_{\rx}(\theta)|^2$ or the magnitude response of AWV at angle $\theta$
\footnote{We experimentally evaluated the coarse beam pattern by categorizing the collected data by \ac{AoA}s with peak RSS, i.e. the estimated \ac{AoA}. The RSS measurements in the $i$th data category then approximate $|\alpha\tilde{\mathbf{w}}^{\hermitian}\mathbf{a}_{\rx}(\theta_k)|^2$ and thus the beam pattern at $\theta_k$, assuming the complex path gain $\alpha$ had constant magnitude throughout the experiment.}, 
of the testbed. A comparison between the measured pattern and the model predicted pattern, $|\mathbf{w}^{\hermitian}\mathbf{a}_{\rx}(\theta)|^2$, is presented in \Cref{fig:result0}, showing an example \ac{DFT} pencil beam and an example \ac{PN} beam\footnote{In the plot, we estimated the magnitude of the complex gain $|\alpha|$ to scale the beam pattern for comparison.}. The results verify the arguments in \Cref{sec:CPR_limitation}. Although the hardware impairment causes little distortion in the mainlobe of \ac{DFT} beam, \ac{DFT} sidelobes and \ac{PN} beam are susceptible to larger distortion.

% Old footnote 5
%\footnote{The beam pattern is defined as the magnitude response of AWV at angle $\theta$, i.e., $|\tilde{\mathbf{w}}^{\hermitian}\mathbf{a}_{\rx}(\theta)|^2$. The coarse pattern measurements in the experiment uses collected data as follows. Firstly, we sorted data based on the indices of peak RSS measurement of DFT sounding. For example, the 1st category contains sounding results under channel realization with approximately the same \ac{AoA} $\theta_1$, which equals to the steering angle of the 1st DFT sounding beam. Hence, the RSS measurements within such data category is approximately $|\alpha\tilde{\mathbf{w}}^{\hermitian}\mathbf{a}_{\rx}(\theta_1)|^2$, which facilitates measuring statistic of beam pattern at $\theta_1$. Note that the path gain $\alpha$ is assumed to have constant magnitude in all LoS realization during experiment.} 

% Old footnote 6
%\footnote{The model predicted beam pattern is defined as $|\mathbf{w}^{\hermitian}\mathbf{a}_{\rx}(\theta)|^2$. In the plot, the magnitude of LoS complex gain $|\alpha|$ is estimated to scaled the beam pattern for comparison.}

\begin{figure*}
     \centering
     \begin{subfigure}[c]{0.28\textwidth}
         \centering
         \includegraphics[width=\textwidth]{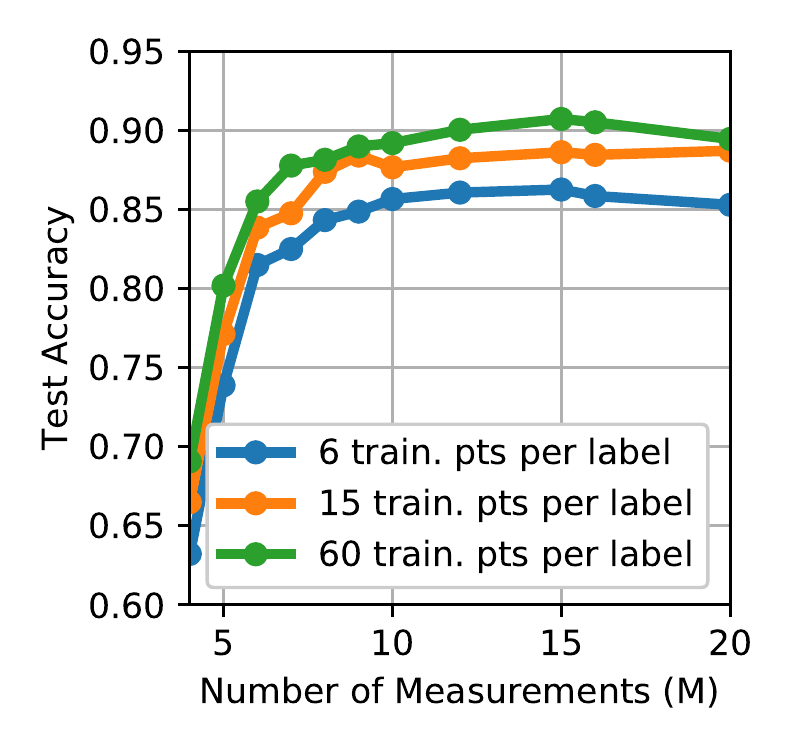}
         \vspace{-5mm}
         \caption{}
         \label{fig:y equals x}
     \end{subfigure}
    %  \hfill
     \begin{subfigure}[c]{0.35\textwidth}
         \centering
         \includegraphics[width=\textwidth]{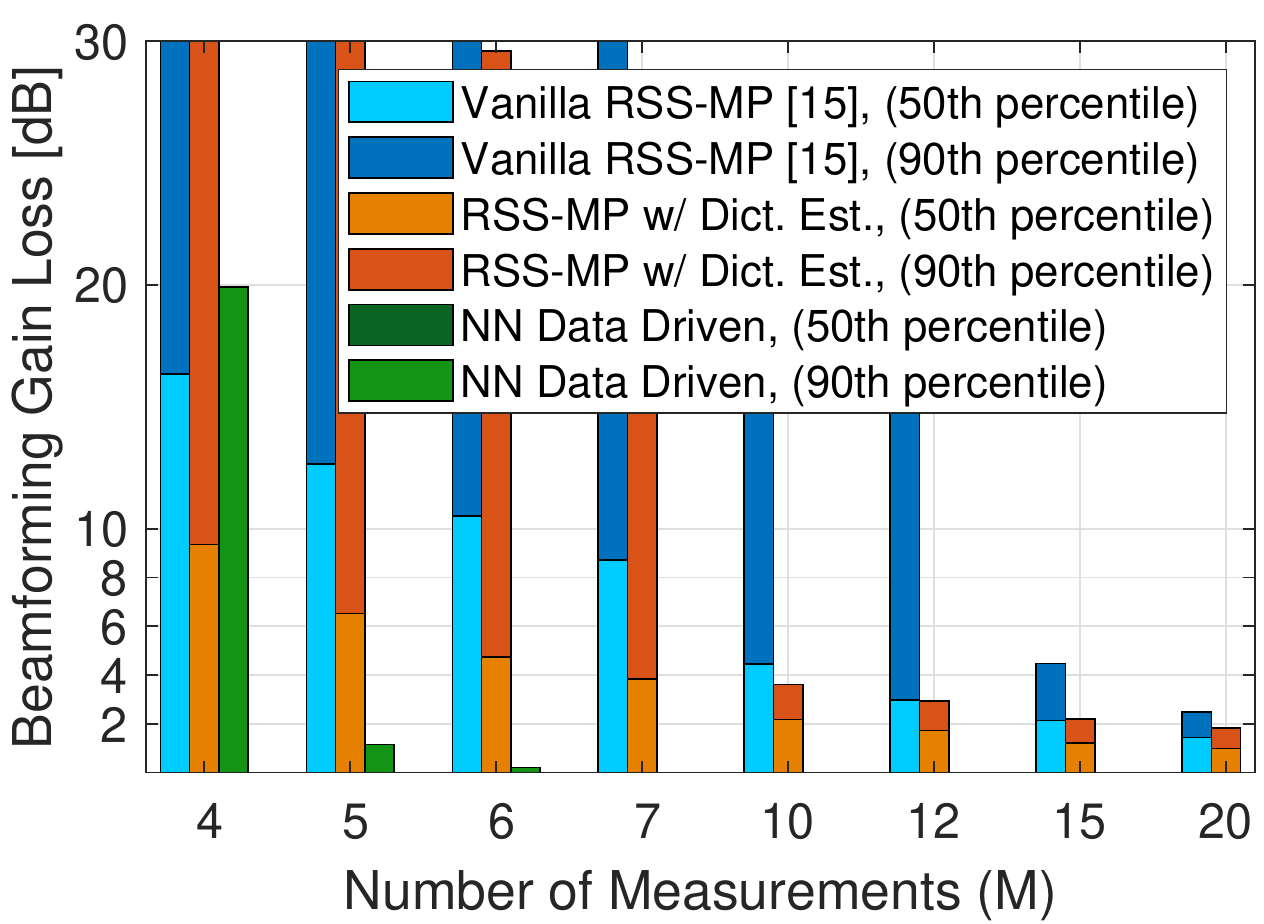}
         \vspace{-5mm}
         \caption{}
         \label{fig:three sin x}
     \end{subfigure}
    %  \hfill
     \begin{subfigure}[c]{0.35\textwidth}
         \centering
         \includegraphics[width=\textwidth]{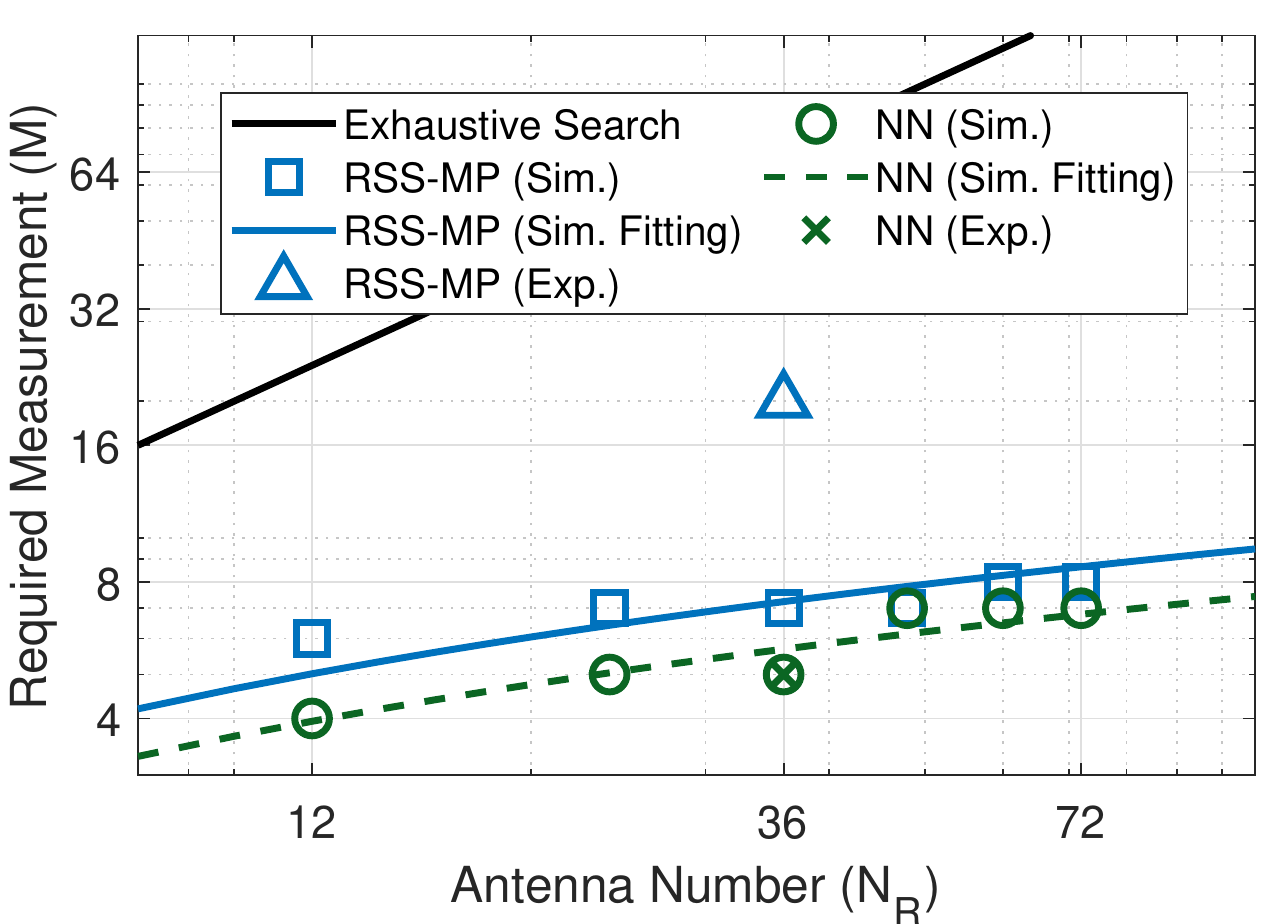}
         \vspace{-5mm}
         \caption{}
         \label{fig:five over x}
     \end{subfigure}
     \vspace{-5mm}
        \caption{(a) Test accuracy as a function of the number of PN beam measurements $M$ for 3 training set sizes. (b) The post-alignment BF gain loss as function of $M$. (c) The required $M$ as function of receiver array size $N_{\rx}$.}
        \label{fig:three graphs}
\end{figure*}

Using the \ac{NN} described in \Cref{sec:neuralnet} with the experimental data, we achieved good accuracy for with a compressed numbers of measurements $M$. Figure \ref{fig:three graphs} (a) shows the test accuracy\footnote{The fraction of DFT beam predictions consistent with the optimal beams from beam sweeps. Incorrect classification may still offer some BF gain.} of the $K=51$ \ac{DFT} beams used for $4 \leq M \leq 20$ and different training set sizes. With $M\geq 6$, the test accuracy saturates around 89\% for the full training set. Performance does improve with more training data, but \ac{NN} is already effective with little training data.
% \color{red}
% Ben, please state the neural network performance in terms of training convergence, evaluation error, or confusion matrix (but not gain loss) with help of \Cref{fig:result1}. 
% \color{black}

% \begin{figure}
% \begin{center}
% \includegraphics[width=0.49\textwidth]{figures/nn_acc_allSNR_exp_trainALL.eps}
% \end{center}
% \vspace{-3mm}
% \caption{Test accuracy as a function of the number of PN beam measurements $M$ for 3 training set sizes.}
% \vspace{-3mm}
% \label{fig:result1}
% \end{figure}

% \begin{figure}
%     \centering
%     \subfigure[]{\includegraphics[width=0.24\textwidth]{figures/nn_acc_allSNR_exp_trainALL.eps}} 
%     \subfigure[]{\includegraphics[width=0.24\textwidth]{monalisa.jpg}} 
%     \subfigure[]{\includegraphics[width=0.24\textwidth]{figures/post_gain_loss_bar.eps}}
%     \subfigure[]{\includegraphics[width=0.24\textwidth]{figures/sim_vs_exp.eps}}
%     \caption{(a) blah (b) blah (c) blah (d) blah}
%     \label{fig:foobar}
% \end{figure}

The post-alignment \ac{BF} gain loss (as compared to a full exhaustive search with $K=51$ measurements) is presented in \Cref{fig:three graphs} (b) with a comparison of three algorithms using \ac{PN} sounding results to predict the best \ac{DFT} beam for data communication. Vanilla \ac{RSS-MP} suffered from the mismatched dictionary information and thus had the poor performance. Even with $M=20$ channel probings, vanilla \ac{RSS-MP} had more than \SI{2}{\decibel} \ac{BF} gain loss in the 90th percentile. With dictionary estimation in RSS-MP, reasonably good alignment is observed with $M\geq 10$. However, precise alignment (less than \SI{2}{\decibel} \ac{BF} gain in 90 percent of test cases) cannot be achieved until $M=20$ measurements. The proposed approach provided further savings, requiring only $M=5$ measurements ($90.2\%$ overhead saving) for comparable post-alignment gain.
% In fact, we observed \SI{0}{\decibel} gain loss for $50$ percent of test cases since the classification accuracy is above 50 percent for $M\geq 4$. 

% \begin{figure}
% \begin{center}
% % \includegraphics[width=0.49\textwidth]{figures/gain_loss.eps}
% \includegraphics[width=0.49\textwidth]{figures/post_gain_loss_bar.eps}
% \end{center}
% \vspace{-3mm}
% \caption{The post-alignment beamforming gain loss as function the number of PN beam probings $M$. The best $50$th percentile and $90$th percentile are shown.}
% \vspace{-3mm}
% \label{fig:result2}
% \end{figure}

In \Cref{fig:three graphs} (c), we compare the required number of measurements\footnote{The minimum $M$ with <2dB BF gain loss compared to a beam sweep in 90\% or more of the predictions} as a function of array size for the \ac{RSS-MP} and proposed algorithms using experimental and simulation data. The simulation used the same \ac{PN} sounding \ac{AWV} realization as in the experiment, but did not include array hardware impairment. The \ac{SNR} in (\ref{eq:pRx_measurement}), i.e., ratio between \ac{RSS} measurements and error variance, is set as \SI{20}{\decibel}. We found the following three observations from the results. Firstly, the required overhead of compressive beam alignment scaled logarithmically with array size, an appealing property for future \ac{mmW} systems. Secondly, the proposed method effectively learned how to solve \ac{CPR} and provided more accurate beam alignment than the heuristic \ac{RSS-MP}, even in the simulations without model mismatch. Lastly, the \ac{NN} had no performance loss in the experimental implementation because the data driven approach is immune to model mismatch due to array imperfection.

% In \Cref{fig:result3}, we use simulation to study the required measurement as a function of array size, and cross compares performance in both simulation and experiment. The simulation uses the same realization of \ac{PN} sounding \ac{AWV} as in the experiment. Moreover, the array hardware impairment is not included in the simulation. We have the following findings. Firstly, the required overhead of compressive beam alignment logarithmic scales with array size which makes it appealing for the future evolution of \ac{mmW} system. Secondly, the comparison between two approaches in the simulation simulation shows that the proposed method effectively learns solving \ac{CPR} and manage to provide more accurate beam alignment than \ac{RSS-MP} even without model mismatch. This is not surprising as RSS-MP is a heuristic algorithm which indicates it is a sub-optimal solution of CPR. Lastly, the implementation loss in \ac{NN} is much lower than \ac{RSS-MP} because the data driven approach is immune to model mismatch due to array imperfection.  

% \begin{figure}
% \begin{center}
% % \includegraphics[width=0.49\textwidth]{figures/gain_loss.eps}
% \includegraphics[width=0.49\textwidth]{figures/sim_vs_exp.eps}
% \end{center}
% \vspace{-3mm}
% \caption{The required number of measurements $M$ for beam alignment to reach <2dB gain loss in the 90th percentile as function of receiver array size.}
% \vspace{-3mm}
% \label{fig:result3}
% \end{figure}
%%%%%%%%%%%%%%%%%%%%%%%%%%%%%%%%%%%
%
%        Conclusions
%
%%%%%%%%%%%%%%%%%%%%%%%%%%%%%%%%%%%
\section{Conclusion and Future works}
\label{sec:Conclusion}
In this work, we presented mmRAPID, a compressive beam alignment scheme that utilizes machine learning to address implementation challenges due to hardware impairments. The results demonstrate that compressive beam alignment can significantly reduce the required number of channel probings. Our implementation on a \SI{60}{\giga\hertz} testbed demonstrates an order of magnitude overhead savings with marginal post-alignment beamforming gain loss, as compared to exhaustive beam sweeps. In the experiment, mmRAPID also outperforms purely model-based compressive methods. %beam alignment.

There are still open questions in this area. The approach and results have yet to be generalized to more sophisticated mmW channels, e.g., non-line-of-sight. Further, the use of compressed channel probing to predict multiple steering directions in multipath environments has yet to be studied. Finally, a comparative study of different sounding codebooks, e.g., multi-lobe beams \cite{Hassanieh:2018:FMW:3230543.3230581,8878130}, with consideration of array impairments and joint design of the codebook and beam alignment algorithm with other machine learning tools are of interest. 

% \color{red} 
% To include more future directions if there is, e.g., Globecom19 best paper on DRL \cite{9014113}.
% \color{black}

% The experiment data and the data processing scripts (Matlab and Python) are accessible via \cite{data_and_code}. \color{red} To update details.
% \color{black}

%%%%%%%%%%%%%%%%%%%%%%%%%%%%%%%%%%%
%
%        Funding agencies
%
%%%%%%%%%%%%%%%%%%%%%%%%%%%%%%%%%%%
\section{Acknowledgments}
This work was supported in part by NSF under grant 1718742. This work was also supported in part by the ComSenTer and CONIX Research Centers, two of six centers in JUMP, a Semiconductor Research Corporation (SRC) program sponsored by DARPA. The \SI{60}{\giga\hertz} Terragraph channel sounders were gifts from Telecom Infra Project (TIP).

\bibliographystyle{ACM-Reference-Format}
\bibliography{sample-base,references,IEEEabrv}


\begin{thebibliography}{22}

%%% the Web address.

\ifx \showCODEN    \undefined \def \showCODEN     #1{\unskip}     \fi
\ifx \showDOI      \undefined \def \showDOI       #1{#1}\fi
\ifx \showISBNx    \undefined \def \showISBNx     #1{\unskip}     \fi
\ifx \showISBNxiii \undefined \def \showISBNxiii  #1{\unskip}     \fi
\ifx \showISSN     \undefined \def \showISSN      #1{\unskip}     \fi
\ifx \showLCCN     \undefined \def \showLCCN      #1{\unskip}     \fi
\ifx \shownote     \undefined \def \shownote      #1{#1}          \fi
\ifx \showarticletitle \undefined \def \showarticletitle #1{#1}   \fi
\ifx \showURL      \undefined \def \showURL       {\relax}        \fi
% invisible to TeX
\providecommand\bibfield[2]{#2}
\providecommand\bibinfo[2]{#2}
\providecommand\natexlab[1]{#1}
\providecommand\showeprint[2][]{arXiv:#2}

\bibitem[\protect\citeauthoryear{{Alkhateeb et al}}{{Alkhateeb et al}}{2018}]%
        {8395149}
\bibfield{author}{\bibinfo{person}{A. {Alkhateeb et al}}.}
  \bibinfo{year}{2018}\natexlab{}.
\newblock \showarticletitle{Deep Learning Coordinated Beamforming for
  Highly-Mobile Millimeter Wave Systems}.
\newblock \bibinfo{journal}{\emph{IEEE Access}}  \bibinfo{volume}{6}
  (\bibinfo{year}{2018}), \bibinfo{pages}{37328--37348}.
\newblock


\bibitem[\protect\citeauthoryear{{Aykin}, {Akgun}, and {Krunz}}{{Aykin}
  et~al\mbox{.}}{2019}]%
        {8878130}
\bibfield{author}{\bibinfo{person}{I. {Aykin}}, \bibinfo{person}{B. {Akgun}},
  {and} \bibinfo{person}{M. {Krunz}}.} \bibinfo{year}{2019}\natexlab{}.
\newblock \showarticletitle{Multi-beam Transmissions for Blockage Resilience
  and Reliability in Millimeter-Wave Systems}.
\newblock \bibinfo{journal}{\emph{{IEEE} J. Sel. Areas Commun.}}
  \bibinfo{volume}{37}, \bibinfo{number}{12} (\bibinfo{year}{2019}),
  \bibinfo{pages}{2772--2785}.
\newblock


\bibitem[\protect\citeauthoryear{{Bajor et al}}{{Bajor et al}}{2019}]%
        {kinget_CS_AOA}
\bibfield{author}{\bibinfo{person}{M. {Bajor et al}}.}
  \bibinfo{year}{2019}\natexlab{}.
\newblock \showarticletitle{A Flexible Phased-Array Architecture for Reception
  and Rapid Direction-of-Arrival Finding Utilizing Pseudo-Random Antenna Weight
  Modulation and Compressive Sampling}.
\newblock \bibinfo{journal}{\emph{{IEEE} J. Solid-State Circuits}}
  \bibinfo{volume}{54}, \bibinfo{number}{5} (\bibinfo{date}{May}
  \bibinfo{year}{2019}), \bibinfo{pages}{1315--1328}.
\newblock


\bibitem[\protect\citeauthoryear{Bertizzolo~et al}{Bertizzolo~et al}{2019}]%
        {10.1145/3349624.3356763}
\bibfield{author}{\bibinfo{person}{L. Bertizzolo~et al}.}
  \bibinfo{year}{2019}\natexlab{}.
\newblock \showarticletitle{MmBAC: Location-Aided MmWave Backhaul Management
  for UAV-Based Aerial Cells}. In \bibinfo{booktitle}{\emph{Proceedings of the
  3rd ACM Workshop on Millimeter-Wave Networks and Sensing Systems}}.
  \bibinfo{pages}{7–12}.
\newblock
\showISBNx{9781450369329}


\bibitem[\protect\citeauthoryear{{Burghal}, {Abbasi}, and {Molisch}}{{Burghal}
  et~al\mbox{.}}{2019}]%
        {9048770}
\bibfield{author}{\bibinfo{person}{D. {Burghal}}, \bibinfo{person}{N.~A.
  {Abbasi}}, {and} \bibinfo{person}{A.~F. {Molisch}}.}
  \bibinfo{year}{2019}\natexlab{}.
\newblock \showarticletitle{A Machine Learning Solution for Beam Tracking in
  mmWave Systems}. In \bibinfo{booktitle}{\emph{2019 53rd Asilomar Conference
  on Signals, Systems, and Computers}}. \bibinfo{pages}{173--177}.
\newblock


\bibitem[\protect\citeauthoryear{{De Donno}, {Palacios}, and {Widmer}}{{De
  Donno} et~al\mbox{.}}{2017}]%
        {7885089}
\bibfield{author}{\bibinfo{person}{D. {De Donno}}, \bibinfo{person}{J.
  {Palacios}}, {and} \bibinfo{person}{J. {Widmer}}.}
  \bibinfo{year}{2017}\natexlab{}.
\newblock \showarticletitle{Millimeter-Wave Beam Training Acceleration Through
  Low-Complexity Hybrid Transceivers}.
\newblock \bibinfo{journal}{\emph{{IEEE} Trans. Wireless Commun.}}
  \bibinfo{volume}{16}, \bibinfo{number}{6} (\bibinfo{year}{2017}),
  \bibinfo{pages}{3646--3660}.
\newblock


\bibitem[\protect\citeauthoryear{et~al}{et~al}{2016}]%
        {7400949}
\bibfield{author}{\bibinfo{person}{R.~W.~Heath et al}.}
  \bibinfo{year}{2016}\natexlab{}.
\newblock \showarticletitle{An Overview of Signal Processing Techniques for
  Millimeter Wave {MIMO} Systems}.
\newblock \bibinfo{journal}{\emph{{IEEE} J. Sel. Topics Signal Process.}}
  \bibinfo{volume}{10}, \bibinfo{number}{3} (\bibinfo{date}{April}
  \bibinfo{year}{2016}), \bibinfo{pages}{436--453}.
\newblock


\bibitem[\protect\citeauthoryear{et~al}{et~al}{2020}]%
        {boljanovic2020truetimedelay}
\bibfield{author}{\bibinfo{person}{V.~Boljanovic et al}.}
  \bibinfo{year}{2020}\natexlab{}.
\newblock \bibinfo{title}{True-Time-Delay Arrays for Fast Beam Training in
  Wideband Millimeter-Wave Systems}.
\newblock
\newblock
\showeprint[arxiv]{eess.SP/2007.08713}


\bibitem[\protect\citeauthoryear{{Ghasempour et al}}{{Ghasempour et
  al}}{2019}]%
        {8871119}
\bibfield{author}{\bibinfo{person}{Y. {Ghasempour et al}}.}
  \bibinfo{year}{2019}\natexlab{}.
\newblock \showarticletitle{Multi-User Multi-Stream mmWave WLANs With Efficient
  Path Discovery and Beam Steering}.
\newblock \bibinfo{journal}{\emph{{IEEE} J. Sel. Areas Commun.}}
  \bibinfo{volume}{37}, \bibinfo{number}{12} (\bibinfo{date}{Dec}
  \bibinfo{year}{2019}), \bibinfo{pages}{2744--2758}.
\newblock


\bibitem[\protect\citeauthoryear{Ghasempour~et al}{Ghasempour~et al}{2020}]%
        {ghasempour2020single}
\bibfield{author}{\bibinfo{person}{Y. Ghasempour~et al}.}
  \bibinfo{year}{2020}\natexlab{}.
\newblock \showarticletitle{Single-shot link discovery for terahertz wireless
  networks}.
\newblock \bibinfo{journal}{\emph{Nature Communications}} \bibinfo{volume}{11},
  \bibinfo{number}{1} (\bibinfo{year}{2020}), \bibinfo{pages}{1--6}.
\newblock


\bibitem[\protect\citeauthoryear{Haider~et al}{Haider~et al}{2018}]%
        {10.1145/3241539.3241542}
\bibfield{author}{\bibinfo{person}{M. Haider~et al}.}
  \bibinfo{year}{2018}\natexlab{}.
\newblock \showarticletitle{LiSteer: MmWave Beam Acquisition and Steering by
  Tracking Indicator LEDs on Wireless APs}. In \bibinfo{booktitle}{\emph{Proc.
  of ACM MobiCom}}. \bibinfo{pages}{273–288}.
\newblock
\showISBNx{9781450359030}


\bibitem[\protect\citeauthoryear{Hassanieh~et al}{Hassanieh~et al}{2018}]%
        {Hassanieh:2018:FMW:3230543.3230581}
\bibfield{author}{\bibinfo{person}{H. Hassanieh~et al}.}
  \bibinfo{year}{2018}\natexlab{}.
\newblock \showarticletitle{Fast Millimeter Wave Beam Alignment}. In
  \bibinfo{booktitle}{\emph{Proc. of ACM SIGCOMM}}. \bibinfo{pages}{432–445}.
\newblock
\showISBNx{978-1-4503-5567-4}


\bibitem[\protect\citeauthoryear{{Heng} and {Andrews}}{{Heng} and
  {Andrews}}{2019}]%
        {9013296}
\bibfield{author}{\bibinfo{person}{Y. {Heng}} {and} \bibinfo{person}{J.~G.
  {Andrews}}.} \bibinfo{year}{2019}\natexlab{}.
\newblock \showarticletitle{Machine Learning-Assisted Beam Alignment for mmWave
  Systems}. In \bibinfo{booktitle}{\emph{Proc. of IEEE GLOBECOM}}.
  \bibinfo{pages}{1--6}.
\newblock


\bibitem[\protect\citeauthoryear{{Nitsche et al}}{{Nitsche et al}}{2015}]%
        {7218630}
\bibfield{author}{\bibinfo{person}{T. {Nitsche et al}}.}
  \bibinfo{year}{2015}\natexlab{}.
\newblock \showarticletitle{Steering with eyes closed: Mm-Wave beam steering
  without in-band measurement}. In \bibinfo{booktitle}{\emph{Proc. of IEEE
  INFOCOM}}. \bibinfo{pages}{2416--2424}.
\newblock


\bibitem[\protect\citeauthoryear{Rasekh and Madhow}{Rasekh and Madhow}{2018}]%
        {Rasekh_2018}
\bibfield{author}{\bibinfo{person}{M. Rasekh} {and} \bibinfo{person}{U.
  Madhow}.} \bibinfo{year}{2018}\natexlab{}.
\newblock \showarticletitle{Noncoherent compressive channel estimation for
  mm-wave massive MIMO}.
\newblock \bibinfo{journal}{\emph{2018 52nd Asilomar Conference on Signals,
  Systems, and Computers}} (\bibinfo{date}{Oct} \bibinfo{year}{2018}).
\newblock
\showISBNx{9781538692189}


\bibitem[\protect\citeauthoryear{Rasekh~et al}{Rasekh~et al}{2017}]%
        {Rasekh_noncoherentCS_ACM_2017}
\bibfield{author}{\bibinfo{person}{M. Rasekh~et al}.}
  \bibinfo{year}{2017}\natexlab{}.
\newblock \showarticletitle{Noncoherent mm{W}ave Path Tracking}. In
  \bibinfo{booktitle}{\emph{Proc. of ACM HotMobile}}. \bibinfo{pages}{13--18}.
\newblock


\bibitem[\protect\citeauthoryear{{Schniter} and {Rangan}}{{Schniter} and
  {Rangan}}{2015}]%
        {6998861}
\bibfield{author}{\bibinfo{person}{P. {Schniter}} {and} \bibinfo{person}{S.
  {Rangan}}.} \bibinfo{year}{2015}\natexlab{}.
\newblock \showarticletitle{Compressive Phase Retrieval via Generalized
  Approximate Message Passing}.
\newblock \bibinfo{journal}{\emph{{IEEE} Trans. Signal Process.}}
  \bibinfo{volume}{63}, \bibinfo{number}{4} (\bibinfo{year}{2015}),
  \bibinfo{pages}{1043--1055}.
\newblock


\bibitem[\protect\citeauthoryear{Sur~et al}{Sur~et al}{2017}]%
        {10.1145/3117811.3117817}
\bibfield{author}{\bibinfo{person}{S. Sur~et al}.}
  \bibinfo{year}{2017}\natexlab{}.
\newblock \showarticletitle{WiFi-Assisted 60 GHz Wireless Networks}. In
  \bibinfo{booktitle}{\emph{Proc. of ACM MobiCom}}. \bibinfo{pages}{28–41}.
\newblock
\showISBNx{9781450349161}


\bibitem[\protect\citeauthoryear{{Va et al}}{{Va et al}}{2019}]%
        {8662770}
\bibfield{author}{\bibinfo{person}{V. {Va et al}}.}
  \bibinfo{year}{2019}\natexlab{}.
\newblock \showarticletitle{Online Learning for Position-Aided Millimeter Wave
  Beam Training}.
\newblock \bibinfo{journal}{\emph{IEEE Access}}  \bibinfo{volume}{7}
  (\bibinfo{year}{2019}), \bibinfo{pages}{30507--30526}.
\newblock


\bibitem[\protect\citeauthoryear{{Wang et al}}{{Wang et al}}{2020}]%
        {9052488}
\bibfield{author}{\bibinfo{person}{Y. {Wang et al}}.}
  \bibinfo{year}{2020}\natexlab{}.
\newblock \showarticletitle{A 39-GHz 64-Element Phased-Array Transceiver With
  Built-In Phase and Amplitude Calibrations for Large-Array 5G NR in 65-nm
  CMOS}.
\newblock \bibinfo{journal}{\emph{{IEEE} J. Solid-State Circuits}}
  \bibinfo{volume}{55}, \bibinfo{number}{5} (\bibinfo{year}{2020}),
  \bibinfo{pages}{1249--1269}.
\newblock


\bibitem[\protect\citeauthoryear{{Yan} and {Cabric}}{{Yan} and
  {Cabric}}{2019}]%
        {8777092}
\bibfield{author}{\bibinfo{person}{H. {Yan}} {and} \bibinfo{person}{D.
  {Cabric}}.} \bibinfo{year}{2019}\natexlab{}.
\newblock \showarticletitle{Compressive Initial Access and Beamforming Training
  for Millimeter-Wave Cellular Systems}.
\newblock \bibinfo{journal}{\emph{{IEEE} J. Sel. Topics Signal Process.}}
  \bibinfo{volume}{13}, \bibinfo{number}{5} (\bibinfo{date}{Sep.}
  \bibinfo{year}{2019}), \bibinfo{pages}{1151--1166}.
\newblock
\showISSN{1941-0484}


\bibitem[\protect\citeauthoryear{Zhang~et al}{Zhang~et al}{2019}]%
        {10.1145/3323679.3326532}
\bibfield{author}{\bibinfo{person}{Y. Zhang~et al}.}
  \bibinfo{year}{2019}\natexlab{}.
\newblock \showarticletitle{Side-Information-Aided Noncoherent Beam Alignment
  Design for Millimeter Wave Systems}. In \bibinfo{booktitle}{\emph{Proc. of
  ACM Mobihoc}}. \bibinfo{pages}{341–350}.
\newblock
\showISBNx{9781450367646}


\end{thebibliography}

\end{document}